\documentclass[12pt]{article}
\usepackage{epsfig}
\usepackage{graphicx}
\usepackage{psfrag}
\usepackage{float}

\begin{document}

\title{Caught Active Walkers}
\author{Steffen Sch\"ollmann\\
Solitudeallee 82,\\
70825 Korntal-M\"unchingen, Germany}

\date{\today\\[2cm]}
\maketitle
\begin{abstract}

\end{abstract}
A discrete implementation on a lattice of the Active Walker Model is presented.
After the model's validity is shown in simple simulations, more complex 
simulations of walkers passing consecutively a lattice from an arbitrary starting point 
at the left border to a random destination on the right border are presented.
It is found that walkers may be caught at a certain position by bouncing back and forth
between two contiguous lattice sites. The statistical
characteristics of this catchment effect are being studied. The probability 
distribution of the number of walkers having passed the lattice before the
catchment occurred shows a exponential decrease to higher numbers. Furthermore
the influence of some parameters of the model on the catchment phenomenon is
discussed. Position and height of the maxima of these distributions show a linear 
dependency on a parameter.

\pagebreak\pagebreak
\setcounter{page}{1}

\section{INTRODUCTION}
Studies of the interaction in many-particle systems has revealed a
wide range of fascinating phenomena. Pattern formation and self-organization 
has been found in chemistry, physics and biology \cite{haken87,haken87a,bak97,paczuski99}.
The same ideas and principles can even be applied in sociology and economics 
\cite{weidlich83,helbing95}. The analytical approach usually models 
a system by differential equations. Thus fluids, reaction-diffusion 
systems and many other systems have been studied. Since several years, 
the enhancing capacity of computers allows to pursue also the 
discrete, numerical approach. Attention has also been paid to 
the simulation of traffic and pedestrian motion \cite{helbing97}. 

Microsimulations of traffic, pedestrian and also granular flow compute 
the motion of each distinct object \cite{wolf96,luding96}. For example, 
the motion of a granular bead is determined by its interaction with the 
environment, i.e. walls, other beads, electrical fields and possibly further 
more. Striking correspondence between simulation and experiment can be found
\cite{schoellmann99}.

\section{Active Walker Model}
The Active Walker Model is based on the social force concept \cite{lewin51,helbing95a}.
According to this idea, environmental influences causing behaviorial 
changes of an individual are modelled by a social force. Thus the change 
in velocity $d\vec{v}/dt$ of a walker is determined by the social force
$\vec{F}_{s}$. $\vec{F}_{s}$ depends not only on the environment 
but also on the preferences and aims of the individual. $\vec{F}_{s}$
is often calculated from a social or environmental potential $V_{\rm env}$.
The walker is called ``active'' because he is able to change the environment
locally. For example, walkers may increase the walking comfort by 
their footprints and thus establish a beaten path \cite{batty97,helbing97a,helbing97b,
helbing98}.

\section{The Discrete Model}
Let the world be a two-dimensional lattice and let time pass in discrete 
steps. At each time $t$, a walker has a certain position $\vec{x}(t)$.
Furthermore, we assume that this walker is on his way to a distinct
destination $\vec{d}$ on our lattice. If reaching his destination
were his one and only will, he would take the direct path along
$\vec{d}-\vec{x}$. But even in a two-dimensional lattice world life may become
more complicated.
Let each lattice site $\vec{x}=(x,y)$ has a set
\begin{equation}
  {\cal N}(\vec{x})=
  \left\{ 
    \begin{array}{l}
      \vec{n}_{1}=(x-1,y),\\ 
      \vec{n}_{2}=(x+1,y),\\
      \vec{n}_{3}=(x,y-1),\\
      \vec{n}_{4}=(x,y+1)
    \end{array}
  \right\}
  \label{equ:set_of_neigh}
\end{equation}
of four neighbouring sites. If $\vec{x}(t)$ is the position of 
the walker at time $t$, $p(\vec{n})$ is the probability that $\vec{n}\in 
{\cal N}(\vec{x}(t))$ will become the next position $\vec{x}(t+1)$ of the 
walker. The neighbouring site with the highest value of $p$ becomes the 
position of the walker 
at time $t+1$. $p(\vec{n})$ is the mathematical expression of the walker's 
aims. If the walker just wants to reach his destination without taking 
the environment into account, the following expression for $p(\vec{n})$ 
would be reasonable:
\begin{equation}
  p(\vec{n})~=~|\vec{x}(t)-\vec{d}| - |\vec{n}-\vec{d}|.
  \label{equ:probab1}
\end{equation}
In general, the influence of the environment can be modelled by a 
environmental potential $V_{\rm env}(\vec{x},t)$. Thus, the probability 
$p(\vec{n})$ becomes
\begin{equation}
  p(\vec{n})~=~w_{\rm dist}\cdot (|\vec{x}(t)-\vec{d}|-|\vec{n}-\vec{d}|)+w_{\rm pot}\cdot V_{\rm env}(\vec{n},t),
  \label{equ:probab2}
\end{equation}
i.e. we calculate the weighted sum of the aim to reach the destination and to 
maximize the environmental potential. The condition $w_{\rm dist}+w_{\rm pot}=1$ has to
be fulfilled.

The environmental potential $V_{\rm env}(\vec{x},t)$ determines how comfortable
the walker feels at site $\vec{x}$ at time $t$. It thus depends heavily on
the individual conditions and aims of the walker. For example, if a walker
runs down a street in order to reach a bus station at the end of this street
and he is also interested in books, he will possibly try to pass book stores
to catch a glimpse. So the environmental potential $V_{\rm env}(\vec{x},t)$ near 
book stores would be higher than near clothes stores (of course!).

In our model, $V_{\rm env}(\vec{x},t)$ reflects the walking comfort $C(\vec{x},t)$
at site $\vec{x}$ at time $t$. If a walker passes a site $\vec{x}$, the 
walking comfort $C(\vec{x},t)$ will increase (because hindering vegetation is being 
damaged). Several walkers running the same
way may cause a beaten path on the long term. On the other hand, 
$C(\vec{x},t)$ decreases by time as beaten paths vanish if they are not used.
Thus, the following differential equation holds for $C(\vec{x},t)$:
\begin{equation}
  \frac{{\rm d}C(\vec{x},t)}{{\rm d}t}~=
    1/\tau\cdot[C_{min}-C(\vec{x},t)] \\
    +~I(\vec{x},t)\cdot[C_{\rm max}-C(\vec{x},t)].
  \label{equ:walk_comf}
\end{equation}
$C_{min}$ is the minimum walking comfort and $C_{\rm max}$ is the maximum one.
$I(\vec{x},t)$ describes the intensity site $\vec{x}$ is frequented by walkers
at time $t$. $1/\tau$ quantifies how fast a beaten path weathers.

The environmental potential $V_{\rm env}(\vec{x},t)$ is mainly but not exclusively
determined by the walking comfort $C(\vec{x},t)$ at site $\vec{x}$. If we used
only $C(\vec{x},t)$, the walker would not be able to recognize a comfortable 
site at some distance and move to it. Clearly this is not realistic. So we
calculate $V_{\rm env}(\vec{x},t)$ as the distance-weighted sum of the walking
comforts of all lattice sites:
\begin{equation}
  V_{\rm env}(\vec{x},t)]~=~\sum_{\vec{y}\neq\vec{x}}e^{-|\vec{y}-\vec{x}|/s(\vec{x},t)}
  \cdot C(\vec{y},t),
  \label{equ:env_pot}
\end{equation}
where $s(\vec{x},t)$ indicates how far one can see at site $\vec{x}$ at time 
$t$.

\section{First Results}

\begin{figure}[H]
  \begin{center}
    \includegraphics[height=5.5cm]{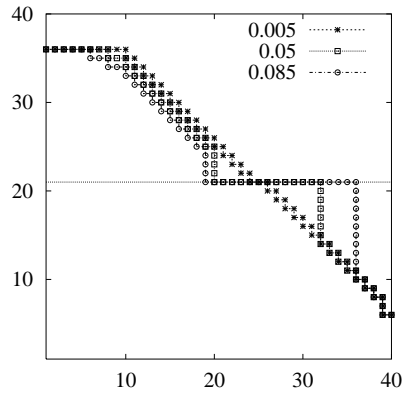}
    \caption[]{A walker crosses a beaten path on the way to its destination. Paths 
resulting from different values for $w_{\rm pot}$ are shown.}
    \label{fig1}
  \end{center}
\end{figure}

Fig. \ref{fig1} illustrates how our model works. We use a $(40,40)$--lattice and
let a walker start at $(1|36)$ with destination $(40|6)$. All lattice sites 
has the same initial walking comfort $C_{0}$ apart from the sites on the 
beaten path in the middle between $(1,21)$ and $(40|21)$ shown as a solid 
line in Fig. \ref{fig1}. If we use $w_{\rm pot}=0.005$ in our discrete active walker 
model, the walker's path is hardly influenced by the environment. The walker looks
for the fastest way to his destination. As Fig. \ref{fig1} reveals, the ``fastest'' 
way on our two-dimensional lattice is not the same as in a continuous two-dimensional 
world. Initially, the walker moves horizontally. Afterwards, he follows a diagonal 
path to his destination. This strange behaviour results from the combination of
restricted movements (horizontal and vertical steps only) and the usage of the 
Euclidian distance in equ. \ref{equ:probab1}. Nevertheless, this effect is not
important for the results presented in this paper. A close look on Fig. \ref{fig1}
shows that the walker follows the beaten path just for 3 steps.

If $w_{\rm pot}$ is increased to $0.05$, the walkers quits its ``direct'' way to the 
destination in order to follow the beaten path. One can also see the effect of
the distance-weighted sum in equ. \ref{equ:env_pot} for $V_{\rm env}$: The walker
is actually attracted by the beaten path. Finally, for $w_{\rm pot}=0.85$ the influence
of the beaten path increases again.

\begin{figure}[H]
  \begin{center}
    \includegraphics[height=5.5cm]{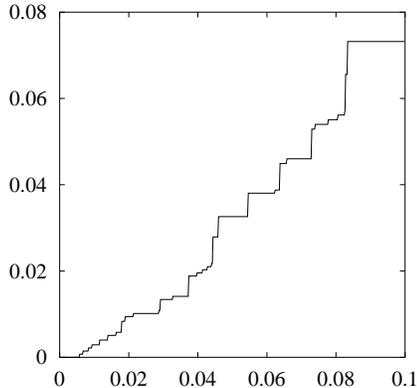}
    \caption[]{$\bar{\Delta}$ for different values of $w_{\rm pot}$ found in the
above simulation (see fig. \ref{fig1}).}
    \label{fig2}
  \end{center}
\end{figure}

We define the mean difference $\bar{\Delta}$ between the actual path of a walker and the 
``direct'' path as
\begin{equation}
  \bar{\Delta}~=~\frac{1}{n}\sum_{\rm i=1}^{n}|\vec{x}_{\rm i}-\vec{x}_{\rm i,dir}|,
  \label{equ:delta}
\end{equation}
where $\vec{x}_{\rm i},~\rm i=1,...,n$ are the $\rm n$ actual positions of the walker on 
its way and
$\vec{x}_{\rm i,dir},~{\rm i=1,...,n}$ are the positions on the direct path to the walker's
destination. $\vec{x}_{\rm i,dir}$ depends on $\vec{x}_{\rm i}$ by equal x values:
$x_{\rm i,dir}=x_{\rm i}$. $\bar{\Delta}$ describes how strong the walker swerved from
the direct way to its destination.
Fig. \ref{fig2} shows the reasonable result that $\bar{\Delta}$ increases if the influence
of the environment becomes higher.

\begin{figure}[H]
  \begin{center}
    \includegraphics[height=7cm]{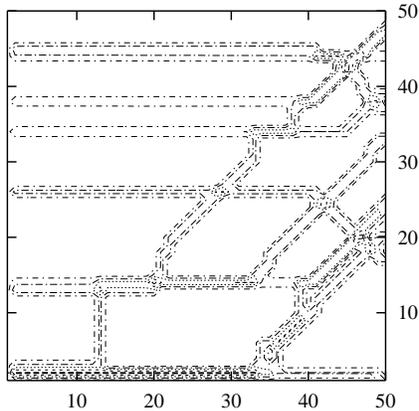}
    \caption[]{Beaten path pattern after 10 walkers have crossed the lattice.
      }
    \label{fig3}
  \end{center}
\end{figure}

Let us do now a more interesting simulation. Initially,
all sites have the same walking comfort. We use a $(50,50)$--lattice. 
A walker starts from a random position on the
left edge to a random position on the right edge. He will definitely use the shortest
way to his destination because there are no beaten paths so far. After the first walker
has reached his destination, we continue with a second walker somewhere on the left border
of the lattice and again a random destination on the right. This walker might be affected
by the beaten path the first one has left on the lattice (see fig. \ref{fig3}). 
My initial interest was to study
the lattice and the walkers after say 1000 iterations:\\
1. Are there persistent beaten paths? How long do they remain?\\
2. What quantity may describe the change of the beaten path pattern from one iteration 
to the next?\\
3. What is the influence of the parameters $s$, $\tau$ and $w_{\rm pot}$?\\

\section{Caught Walkers}

At some point in my studies, I wondered why a walker seems to never reach its destination.
I had a closer look and found out that he was bouncing from one site to a contiguous one
and vice versa all the time. First, I thought of an error in the implementation of the 
model. But I could not find any. So I checked the algorithm by calculating $p(\vec{n})$ 
by hand - and astonishingly, my program was correct.The explanation is simple. 
In direction of his destination the sites offer 
much less walking comfort than the site he has been before. Although this artefact 
obviously diminishes the applicability of the model, it is worth a close look.

It was obvious that the number of walkers having reached their destinations successfully 
before a walker is being caught differed from one simulation run to the next. Thus
I was interested in the probability distribution $p(n)$ of the number of walkers $n$ 
reaching their destination. Further on I investigated how this probability distribution 
$p(n)$ depended on the parameters of the model, i.e. $w_{\rm pot}$, $\tau$ or even the 
size of the lattice. The distributions $p(n)$ shown below are normalized:
\begin{equation}
  \int_{0}^{n_{\rm max}}p(n){\rm dn}~\approx~\sum_{n=0}^{n_{\rm max}} p(n)\cdot \Delta n ~=~ 1.
  \label{equ:normalization1d}
\end{equation}
$\Delta n=1$ is the bin width of the discrete probability distribution and 
$n_{\rm max}$ is the maximum number of walkers.

\begin{figure}[H]
  \begin{center}
    (a)\includegraphics[height=5.3cm]{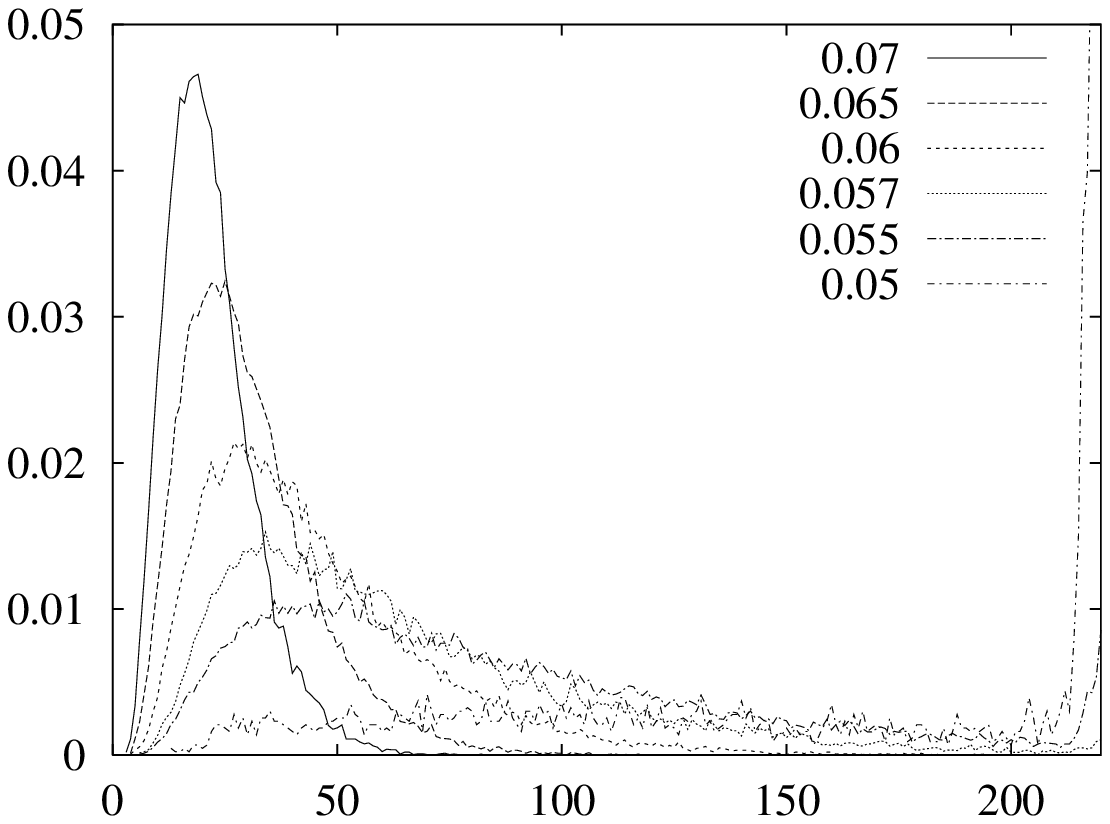}
    (b)\includegraphics[height=5.3cm]{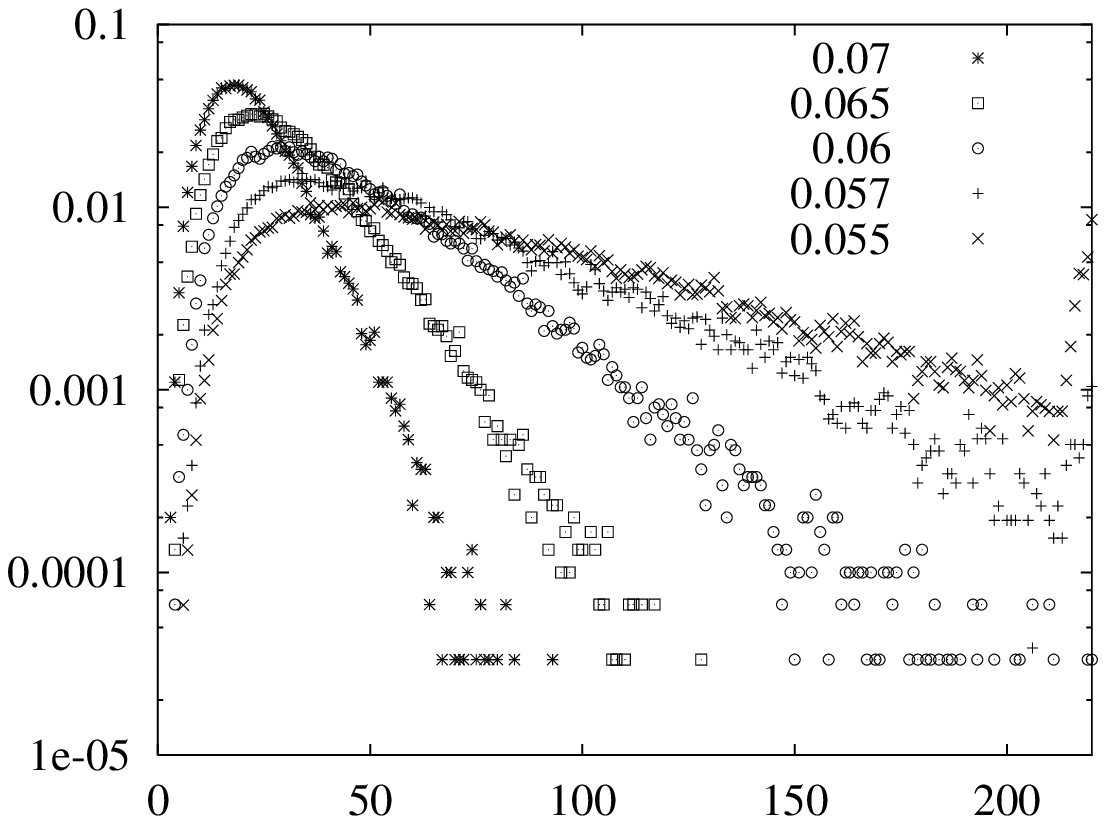}
    \caption[]{$p(n)$ with $n_{\rm max}=220$ for different values of $w_{\rm pot}$ 
with linear (a) and logarithmic (b) y-axis.}
    \label{fig4}
  \end{center}
\end{figure}

Fig. \ref{fig4} elucidates the dependency of $p(n)$ on $w_{\rm pot}$. 
Generally speaking, if the influence of the environment and thus $w_{\rm pot}$ is 
increased, the walkers are being caught faster. $p(n)$ shows a distinct maximum. The plots 
for $w_{\rm pot}=0.055$ and  $0.5$ show a peak at $n=220$. Because the simulation was 
stopped after 220 walkers had reached their destinations, the probability $p(n\geq 220)$ is 
aggregated in $p(220)$. Only for small values of $w_{\rm pot}$ $p(n\geq 220)$ is neglectable.

The half-logarithmic plot in fig. \ref{fig4}a shows that $p(n)$ decreases exponentially 
for increasing $n$.

\begin{figure}[H]
  \begin{center}
    \includegraphics[height=6.5cm]{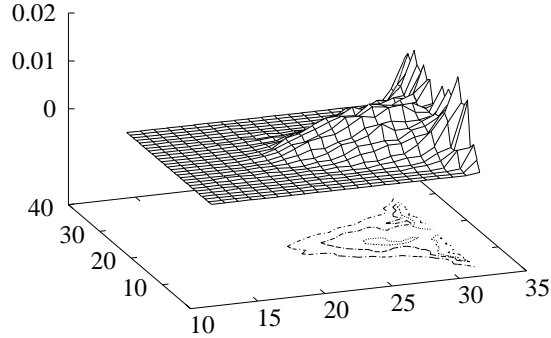}
    \caption[]{Surface and contour plot of $p(\vec{x})$ for $w_{\rm pot}=0.06$}.
    \label{fig5}
  \end{center}
\end{figure}

Where are the ``prisons'' of the walkers located on the lattice? In order to answer
this question, I determined the normalized probability distribution $p(\vec{x})$ for the
locations $\vec{x}$ of the prisons. $p(\vec{x}=(x|y))$ fulfills the equation
\begin{equation}
  \int_{x} \int_{y} p(x,y){\rm dxdy}~\approx~\sum_{x=0}^{x_{\rm max}} \sum_{y=0}^{y_{\rm max}} p(x,y)\cdot \Delta x \Delta y ~=~ 1,
  \label{equ:normalization2d}
\end{equation}
where $x_{\rm max}=y_{\rm max}=40$ is the size of the lattice and $\Delta x=\Delta y=1$ the
size of the bins.

Fig. \ref{fig5} shows that $p(\vec{x})$ is nearly symmetrical to $y=20$. Please note 
that for $x<12$ and $x>32$ no prisons are found at all. The highest values for 
$p(\vec{x})$ are found for $x=32$. Astonishingly, walkers are not being caught next to 
the right border of the lattice. 

\begin{figure}[H]
  \begin{center}
    (a)\includegraphics[height=5.3cm]{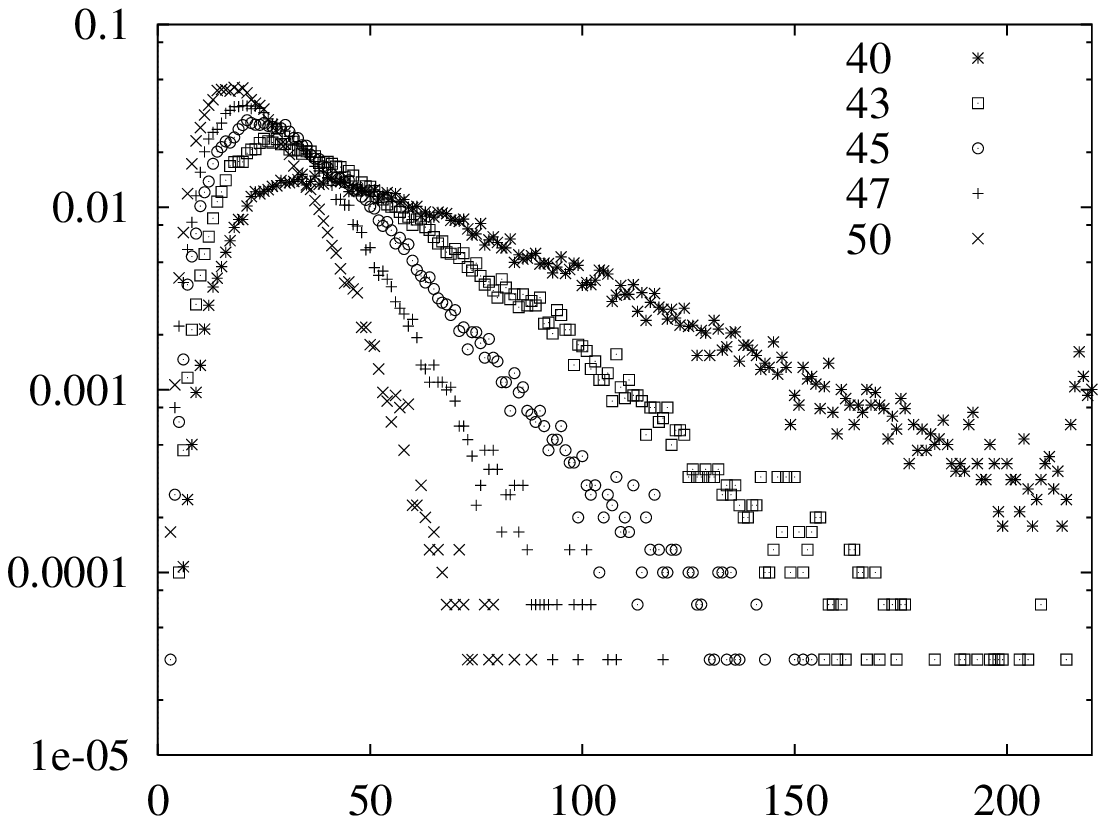}
    (b)\includegraphics[height=5.3cm]{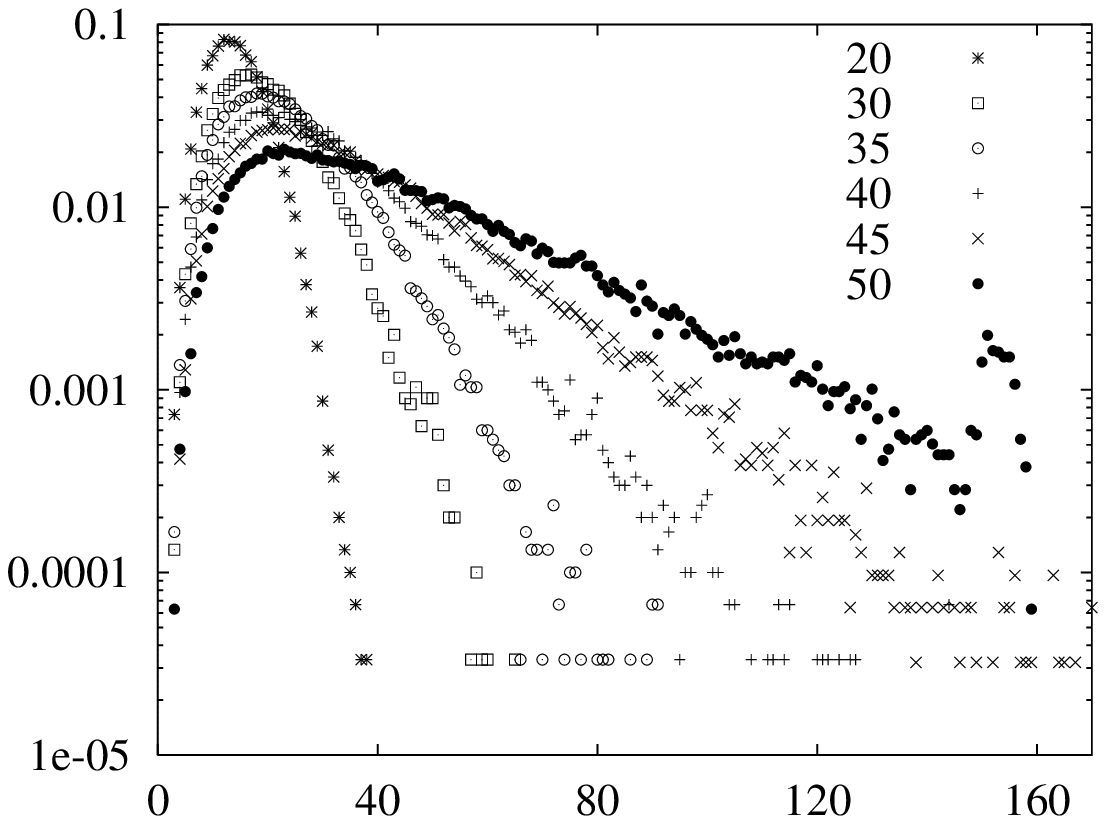}
    \caption[]{Dependency of $p(n)$ on $\tau$ (a) and the lattice size
(b). The labels in (a) are the values for $\tau$ and in (b) the number of sites in each 
dimension. $n_{\rm max}=220$ in (a) and $180$ in (b).}
    \label{fig6}
  \end{center}
\end{figure}

The probability distributions $p(n)$ for different $\tau$ and
different lattice sizes are shown in fig. \ref{fig6} in half-logarithmic plots. They are 
very similar to the curves discussed above and also show exponential decay for increasing $n$.

\begin{figure}[H]
  \begin{center}
    (a)\includegraphics[height=5.3cm]{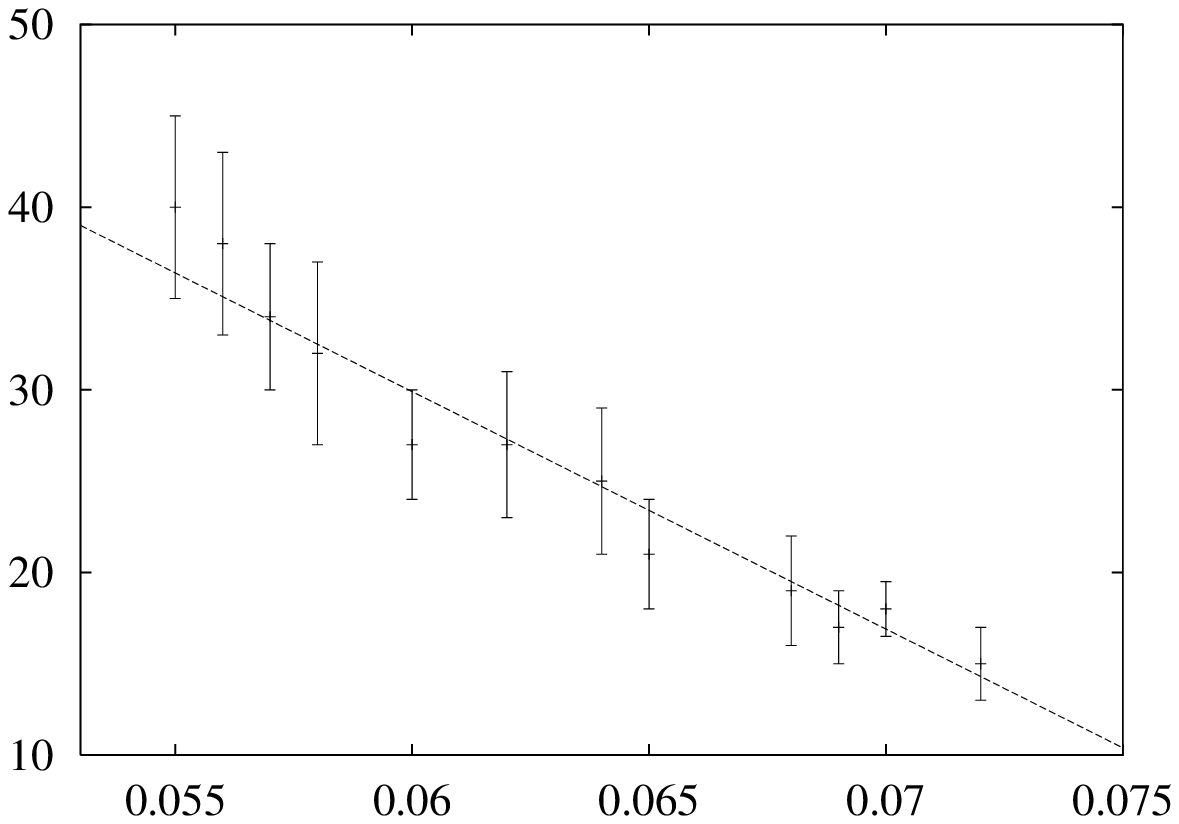}
    (b)\includegraphics[height=5.3cm]{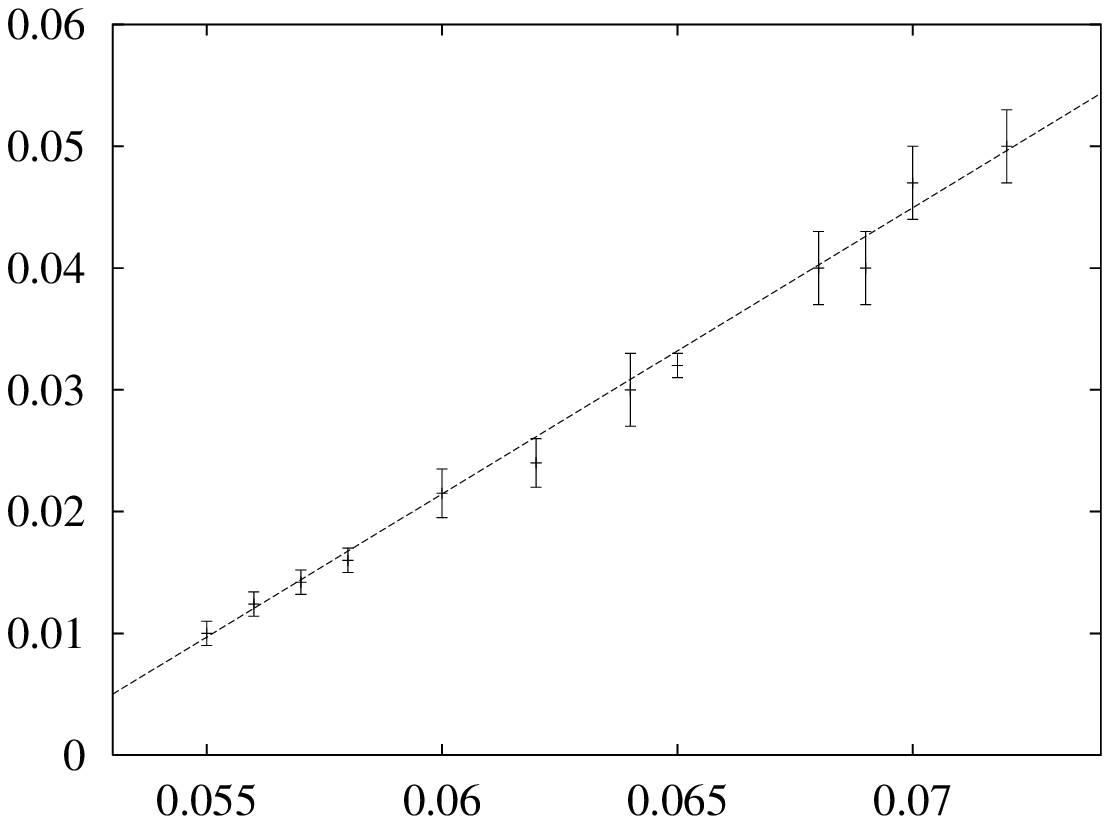}
    \caption[]{Position $n_{\rm max}$ (a) and height $p_{\rm max}$ (b) of the maximum of $p(n)$ 
for different values of $w_{\rm pot}$. The linear fits $y=a+m\cdot x$ are done with $a=114$ and $m=-1403$ in (a) and $a=-0.1199$ and $m=2.349$ in (b).}
    \label{fig7}
  \end{center}
\end{figure}

Now we have a closer look on the maxima of the plots in fig. \ref{fig4}a. The maxima are found
at different positions $n_{\rm max}$ and have different heights $p_{\rm max}$. 
Fig. \ref{fig7} shows $n_{\rm max}$ and $p_{\rm max}$ for different values of $w_{\rm pot}$.
Obviously, $n_{\rm max}$ decreases for higher $w_{\rm pot}$ whereas the maxima becomes 
higher. Especially $p_{\rm max}$ shows nearly a linear dependency on $w_{\rm pot}$.

\section{Outlook}
The described ``catchment'' phenomenon is an artefact of the introduced discrete 
active walker model. Thus the model has to be improved in order to allow real-life
simulations. On the other hand, it would be interesting to compare the described statistical 
characteristics of the catchment phenomenon with those of real-world catchment effects for 
example in particle physics.

----------------------------------------------------------------------

\nopagebreak
\bibliographystyle{unsrt}
\bibliography{caught_walker}

\end{document}